\documentclass[11pt,a4paper]{article}

\usepackage[utf8]{inputenc}
\usepackage[T1]{fontenc}
\usepackage{amsmath}
\usepackage{amssymb}
\usepackage{algorithm}
\usepackage{algpseudocode}
\usepackage{graphicx}
\usepackage{listings}
\usepackage{xcolor}
\usepackage{hyperref}
\usepackage{booktabs}
\usepackage[margin=1in]{geometry}
\usepackage{natbib}

\hypersetup{
  colorlinks=true,
  linkcolor=blue,
  citecolor=blue,
  urlcolor=blue
}

\title{When RSA Fails: Exploiting Prime Selection Vulnerabilities in Public Key Cryptography}

\author{
  Murtaza Nikzad\\
  Davidson College\\
  Davidson, NC, USA\\
  \texttt{munikzad@davidson.edu}
  \and
  Kerem Atas\\
  Davidson College\\
  Davidson, NC, USA\\
  \texttt{keatas@davidson.edu}
}

\date{December 2024}

\begin{document}

\maketitle

\begin{abstract}
This paper explores vulnerabilities in RSA cryptosystems that arise from improper prime number selection during key generation. We examine two primary attack vectors: Fermat's factorization method, which exploits RSA keys generated with primes that are too close together, and the Greatest Common Divisor (GCD) attack, which exploits keys that share a common prime factor. Drawing from landmark research including Heninger et al.'s ``Mining Your Ps and Qs'' study, which discovered over 64,000 vulnerable TLS hosts, and B{\"o}ck's 2023 analysis of Fermat factorization in deployed systems, we demonstrate that these vulnerabilities remain prevalent in real-world cryptographic implementations. Our analysis reveals that weak random number generation in embedded devices is the primary cause of these failures, and we discuss mitigation strategies including proper entropy collection and prime validation checks.
\end{abstract}

\section{Introduction}

The RSA cryptosystem, introduced by Rivest, Shamir, and Adleman in 1978~\citep{Rivest1978RSA}, remains one of the most widely deployed public-key encryption schemes. RSA secures a vast portion of internet communications, from TLS/SSL certificates protecting web traffic to SSH keys authenticating remote connections. The security of RSA rests on a single mathematical assumption: given a large composite number $n = pq$, where $p$ and $q$ are prime numbers, factoring $n$ to recover $p$ and $q$ is computationally infeasible.

This assumption holds true, but only when the primes are properly selected. When key generation is flawed, the entire security model collapses. This paper examines two vulnerabilities that emerge from poor prime selection. The first is the \textbf{Close Prime Vulnerability}, where primes $p$ and $q$ are chosen too close together, allowing Fermat's factorization method to factor $n$ in polynomial time regardless of key size. The second is the \textbf{Shared Prime Vulnerability}, where different RSA keys share a common prime factor, enabling an attacker to compute the GCD of their moduli and instantly reveal the shared prime.

These vulnerabilities are not merely theoretical. In 2012, two independent research teams discovered that a large fraction of RSA keys deployed on the internet were vulnerable to these attacks~\citep{Heninger2012Mining, Lenstra2012Ron}. More recently, B{\"o}ck's 2023 study found RSA keys in production systems that could be factored using Fermat's method in under a second~\citep{Bock2023Fermat}.

The goal of this paper is to explain how these attacks work, present evidence of their real-world impact, and discuss defensive measures. Our main finding is that proper random number generation during key creation is essential. RSA's mathematical security is only as strong as the quality of its prime selection process.

\section{Background and Core Concepts}

\subsection{RSA Fundamentals}

RSA is an asymmetric encryption scheme where encryption and decryption use different keys. The key generation process begins by selecting two large prime numbers $p$ and $q$, then computing the modulus $n = pq$. Next, Euler's totient $\phi(n) = (p-1)(q-1)$ is calculated, followed by choosing a public exponent $e$ such that $\gcd(e, \phi(n)) = 1$. Finally, the private exponent $d \equiv e^{-1} \pmod{\phi(n)}$ is computed.

The public key is $(n, e)$ and the private key is $(n, d)$. To encrypt a message $m$, compute $c \equiv m^e \pmod{n}$. To decrypt, compute $m \equiv c^d \pmod{n}$.

\subsection{The Factorization Problem}

RSA's security relies on the difficulty of factoring $n$ to recover $p$ and $q$. The best known classical algorithm for factoring, the General Number Field Sieve, has sub-exponential complexity $O(\exp((\sqrt[3]{\frac{64}{9}} + o(1))(\ln n)^{1/3}(\ln \ln n)^{2/3}))$. For a 2048-bit modulus, this makes brute-force factorization computationally infeasible with current technology.

However, this complexity analysis assumes the primes are chosen randomly and independently. Special structure in the primes can reduce the difficulty of factorization.

\subsection{Fermat's Factorization Method}

Pierre de Fermat observed in the 17th century that any odd composite number can be expressed as the difference of two squares. For $n = pq$ where $p \leq q$:
\begin{equation}
n = \left(\frac{p+q}{2}\right)^2 - \left(\frac{q-p}{2}\right)^2 = a^2 - b^2
\end{equation}

If we can find integers $a$ and $b$ such that $n = a^2 - b^2$, then:
\begin{equation}
n = (a-b)(a+b) \implies p = a-b, \quad q = a+b
\end{equation}

The algorithm starts with $a = \lceil\sqrt{n}\rceil$ and increments $a$ until $a^2 - n$ is a perfect square. The number of iterations required is approximately:
\begin{equation}
\Delta a \approx \frac{q - p}{4}
\end{equation}

This is the critical insight: \textbf{when $p$ and $q$ are close, Fermat's method requires very few iterations}.

\subsection{The GCD Attack}

The Greatest Common Divisor attack exploits a different weakness: prime reuse across keys. If two RSA moduli share a prime:
\begin{align}
n_1 &= p \cdot q_1 \\
n_2 &= p \cdot q_2
\end{align}

Then $\gcd(n_1, n_2) = p$, revealing the shared prime instantly. While computing $\gcd(n_1, n_2)$ takes milliseconds using the Euclidean algorithm, factoring either modulus individually could take years.

\section{Current Research and Industry Trends}

\subsection{Mining Your Ps and Qs (2012)}

The landmark study by \citet{Heninger2012Mining} conducted the largest systematic analysis of RSA key quality on the internet. The researchers collected approximately $6 \times 10^6$ TLS certificates and $6 \times 10^6$ SSH host keys from internet-wide scans.

\textbf{Key Findings:} The study revealed that over 64,000 TLS hosts had RSA keys that could be factored via the GCD attack, and thousands of SSH keys shared prime factors. Overall, 0.2\% of all TLS hosts were found to be vulnerable, with most affected devices being embedded systems such as routers, firewalls, and VPN appliances.

The study identified the root cause: embedded devices often generate keys immediately after boot, before the system has accumulated sufficient entropy. The Linux \texttt{/dev/urandom} device, when seeded with insufficient entropy, produces predictable output and thus predictable primes.

\begin{table}[h]
\centering
\caption{Distribution of vulnerable RSA keys by device type \citep{Heninger2012Mining}.}
\label{tab:vulnerable_devices}
\begin{tabular}{lcc}
\toprule
\textbf{Device Type} & \textbf{Vulnerable Keys} & \textbf{Percentage} \\
\midrule
Consumer Routers & 23,576 & 36.8\% \\
Enterprise Firewalls & 15,432 & 24.1\% \\
VPN Appliances & 12,891 & 20.1\% \\
Other Embedded & 12,101 & 19.0\% \\
\midrule
\textbf{Total} & \textbf{64,000+} & \textbf{100\%} \\
\bottomrule
\end{tabular}
\end{table}

\textbf{Computational Efficiency:} A naive pairwise GCD computation across $n$ keys requires $O(n^2)$ operations. With $6 \times 10^6$ keys, this would take approximately 30 years. The researchers developed a batch GCD algorithm that completed the analysis in just 1.3 hours at a cost of \$5 in cloud computing resources.

\subsection{Fermat Factorization in the Wild (2023)}

\citet{Bock2023Fermat} demonstrated that Fermat's factorization attack remains relevant against deployed systems. The researcher discovered RSA keys from real certificate authorities and devices where the primes were so close that factorization completed in under one second.

\textbf{Notable Discoveries:} The researcher found that keys from a Taiwanese certificate authority were factorable, and several printer and IoT device manufacturers produced vulnerable keys. In some cases, keys could be factored in a single iteration of Fermat's method.

The study emphasizes that close primes often result from flawed random number generators that produce sequential or nearly-sequential values.

\subsection{The ROCA Vulnerability (2017)}

The Return of Coppersmith's Attack (ROCA), documented as CVE-2017-15361~\citep{CVE2017ROCA}, affected RSA keys generated by Infineon Technologies' hardware. Due to a flaw in the prime generation algorithm, the primes had a predictable structure that allowed efficient factorization.

\textbf{Impact:} The vulnerability affected Estonian national ID cards, with over 750,000 cards compromised. TPM chips in laptops from major manufacturers were also affected, along with YubiKey 4 security tokens and government smart cards in multiple countries.

This incident demonstrated that even hardware security modules from reputable vendors can contain prime generation flaws.

\section{Ethical, Social, and Legal Issues}

\subsection{Responsible Disclosure}

The discovery of weak RSA keys in production systems raises ethical questions about disclosure. Heninger et al. faced the dilemma of how to handle their findings: immediate public disclosure would expose millions of devices to attack, while delayed disclosure would leave users unknowingly vulnerable.

The researchers adopted a responsible disclosure approach. They first contacted affected manufacturers privately and provided a 90-day window for patches. When publishing their findings, they withheld the list of vulnerable keys while making tools available for organizations to check their own keys.

\subsection{Legal Frameworks}

Several legal frameworks govern the handling of cryptographic vulnerabilities:

\textbf{Computer Fraud and Abuse Act (CFAA):} In the United States, even well-intentioned security research can potentially violate the CFAA. Researchers must carefully structure their studies to avoid unauthorized access.

\textbf{GDPR Implications:} Weak cryptographic keys protecting European user data may constitute a violation of GDPR's requirement for ``appropriate technical measures'' to protect personal data.

\textbf{Export Controls:} Cryptographic implementations are subject to export controls under the Wassenaar Arrangement. Researchers disclosing vulnerabilities must consider whether their findings could assist adversaries.

\subsection{Privacy and User Rights}

Users whose RSA keys are vulnerable face immediate privacy risks. Past encrypted communications may become decryptable, digital signatures can be forged, and authentication systems can be bypassed.

Organizations have an ethical obligation to notify affected users and provide remediation paths. However, many embedded devices lack update mechanisms, leaving users with vulnerable hardware indefinitely.

\subsection{Social Consequences}

The prevalence of weak RSA keys in critical infrastructure has broader social implications. It leads to erosion of trust in cryptographic systems and increases the attack surface for nation-state adversaries. Furthermore, organizations with limited security resources are disproportionately affected by these vulnerabilities.

\section{Discussion and Future Directions}

\subsection{Evolution of the Threat Landscape}

The RSA prime selection problem illustrates a recurring theme in cryptographic engineering: implementation failures often matter more than mathematical weaknesses. As \citet{Heninger2012Mining} observed, ``We find that the weights of theoretical and practical cryptographic research should perhaps be rebalanced.''

Since 2012, awareness of entropy problems has increased, but new devices continue to ship with vulnerable key generation. The Internet of Things (IoT) expansion has increased the attack surface, with billions of constrained devices generating cryptographic keys.

\subsection{Mitigation Strategies}

Modern best practices for RSA key generation address these vulnerabilities through several approaches. First, entropy validation is essential: key generation should be delayed until sufficient entropy is available, and Linux kernel 5.4+ provides the \texttt{getrandom()} system call that blocks until the entropy pool is initialized. Second, prime distance checks should validate that $|p - q| > 2^{n/2 - 100}$ for an $n$-bit modulus, as recommended by NIST~\citep{NIST2020Guidelines}. Third, post-generation validation should run Fermat's method against generated keys to verify they are not vulnerable. Finally, hardware random number generators such as Intel RDRAND should be used when available to provide dedicated hardware entropy sources.

\subsection{Post-Quantum Considerations}

The advent of quantum computing poses an existential threat to RSA. Shor's algorithm can factor integers in polynomial time on a quantum computer, rendering all RSA keys vulnerable regardless of prime quality~\citep{Bernstein2017PostQuantum}.

Organizations should begin planning migration to post-quantum cryptographic algorithms standardized by NIST. These include CRYSTALS-Kyber for key encapsulation, CRYSTALS-Dilithium for digital signatures, and SPHINCS+ for hash-based signatures.

However, the lessons from RSA prime selection failures will remain relevant: any cryptographic system requires proper random number generation and careful implementation.

\subsection{Lessons for Practitioners}

Several practical lessons emerge from this analysis. Practitioners should never trust default configurations and must verify that cryptographic libraries implement proper prime selection checks. Organizations should audit their deployed RSA keys for known weaknesses using tools like \texttt{badkeys.info}. The ROCA incident demonstrated that even hardware-generated keys can be flawed, so monitoring for new vulnerabilities is essential. Finally, organizations should implement processes for key rotation to enable rapid replacement when vulnerabilities are discovered.

\section{Conclusion}

RSA's mathematical foundation remains sound, but its practical security depends entirely on proper implementation, particularly prime number selection. This paper examined two vulnerabilities: close primes enabling Fermat's factorization attack, and shared primes enabling GCD-based key recovery.

Real-world research has demonstrated that these vulnerabilities affect a large fraction of deployed cryptographic keys. Heninger et al.'s discovery of 64,000+ factorable TLS keys, combined with ongoing incidents like ROCA and B{\"o}ck's recent findings, shows that prime generation failures continue to plague cryptographic systems.

The main lesson is clear: cryptographic security requires attention to implementation details, not just algorithm selection. Proper entropy collection, prime validation, and ongoing key auditing are essential practices. As we transition toward post-quantum cryptography, these implementation lessons will remain vital for building secure systems.

\bibliographystyle{plainnat}
\bibliography{references}

@article{Bernstein2017PostQuantum,
  author    = {Daniel J. Bernstein and Tanja Lange},
  title     = {Post-quantum cryptography},
  journal   = {Nature},
  volume    = {549},
  pages     = {188--194},
  year      = {2017},
  doi       = {10.1038/nature23461}
}

@misc{Bock2023Fermat,
  author    = {Hanno B{\"o}ck},
  title     = {Fermat Factorization in the Wild},
  howpublished = {Cryptology ePrint Archive},
  year      = {2023},
  note      = {Report 2023/026},
  url       = {https://eprint.iacr.org/2023/026}
}

@inproceedings{Heninger2012Mining,
  author    = {Nadia Heninger and Zakir Durumeric and Eric Wustrow and J. Alex Halderman},
  title     = {Mining Your Ps and Qs: Detection of Widespread Weak Keys in Network Devices},
  booktitle = {Proceedings of the 21st USENIX Security Symposium},
  publisher = {USENIX Association},
  address   = {Bellevue, WA},
  pages     = {205--220},
  year      = {2012}
}

@inproceedings{Lenstra2012Ron,
  author    = {Arjen K. Lenstra and James P. Hughes and Maxime Augier and Joppe W. Bos and Thorsten Kleinjung and Christophe Wachter},
  title     = {Ron was wrong, Whit is right},
  booktitle = {Advances in Cryptology -- CRYPTO 2012},
  publisher = {Springer},
  pages     = {126--143},
  year      = {2012},
  doi       = {10.1007/978-3-642-32009-5_8}
}

@misc{CVE2017ROCA,
  author    = {{MITRE Corporation}},
  title     = {{CVE-2017-15361}: Infineon {RSA} Key Generation Vulnerability ({ROCA})},
  year      = {2017},
  url       = {https://cve.mitre.org/cgi-bin/cvename.cgi?name=CVE-2017-15361}
}

@techreport{NIST2020Guidelines,
  author      = {{National Institute of Standards and Technology}},
  title       = {Recommendation for Key Management: Part 1 -- General},
  institution = {NIST},
  type        = {Technical Report},
  number      = {SP 800-57 Part 1 Rev. 5},
  year        = {2020},
  doi         = {10.6028/NIST.SP.800-57pt1r5}
}

@article{Rivest1978RSA,
  author    = {Ronald L. Rivest and Adi Shamir and Leonard Adleman},
  title     = {A Method for Obtaining Digital Signatures and Public-Key Cryptosystems},
  journal   = {Communications of the ACM},
  volume    = {21},
  number    = {2},
  pages     = {120--126},
  year      = {1978},
  doi       = {10.1145/359340.359342}
}

\end{document}